# Wireless Access to Ultimate Virtual Reality 360-Degree Video at Home

Huanle Zhang, Ahmed Elmokashfi, Zhicheng Yang, and Prasant Mohapatra

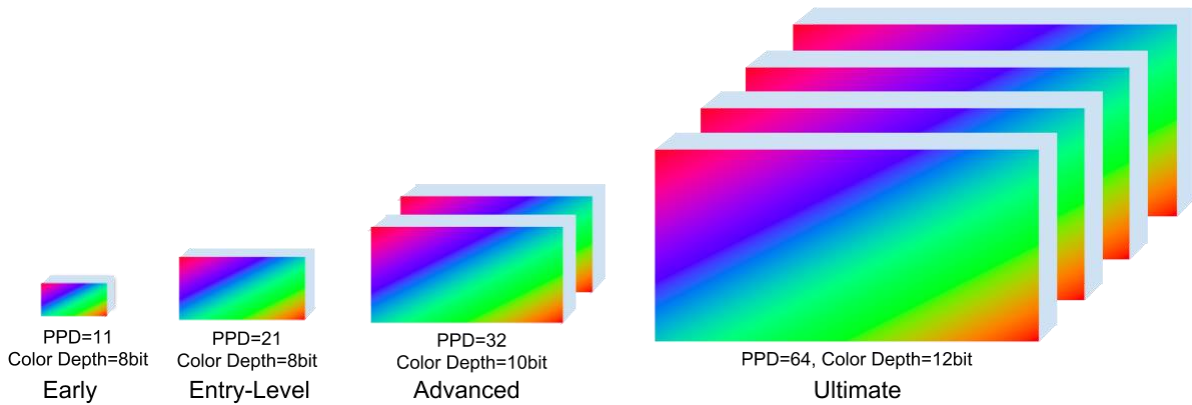

Fig. 1. Illustration of VR 360 video size in the early stage, the entry-level experience stage, the advanced experience stage and the ultimate experience stage [1]. Each block represents 30 frames per second. As of early 2019, only the early stage is fully supported

**Abstract**—VR 360 videos are data-hungry and latency-sensitive that pose unique challenges to the networking infrastructure. In this paper, we focus on the ultimate VR 360 that satisfies human eye fidelity. The ultimate VR 360 requires downlink of 1.5 Gbps for viewing and uplink of 6.6 Gbps for live broadcasting, with round-trip time of less than 8.3 ms. On the other hand, wireless access to VR 360 services is preferred over wireline transmission because of the better user experience and safety concerns (e.g., tripping hazard). We explore in this paper whether available and upcoming cellular communications and WiFi communications can support the ultimate VR 360 at home. Specifically, we consider 5G, IEEE 802.11ac and IEEE 802.11ad. According to their specifications and empirical measurements, we find that: (1) 5G cannot support the ultimate VR 360 because of the limited user data rate; (2) IEEE 802.11ac can support a single-user ultimate VR 360 viewing, but fails to support the live broadcasting; (3) IEEE 802.11ad has the potential to support a single-user ultimate VR 360 viewing with implementation enhancements to reduce the latency. None of the existing wireless technologies can fully support multiple ultimate VR 360 users at home. Our findings highlight the need for developing more advanced wireless technologies.[1]

## I. Introduction

As the next-generation Internet platform, Virtual Reality (VR) has aroused great interest in both academia and industry. The global VR market was valued at USD 3.13 billion in 2017 and is expected to reach USD 49.7 billion by 2023, at a compound annual growth rate of 58.54% over 2018-2023 [2]. VR systems are expected to ultimately support human perceptions such as sight, hearing, touch, smell and taste. Currently, academia and industry are focused on sight and hearing, in the form of VR videos. A VR video system can be further classified into VR 360 (immersive or panoramic videos), free view-point, computer graphics.

VR 360 is expected to be the first popular online VR application [1]. It has two forms of use: viewing and live broadcasting. In VR 360 viewing, a user wears a VR headset that blocks outside view so that the user immerses only on what is being displayed on the headset. The user watches a part of the view, termed as the Field-of-View (FoV), while the full-view video is captured 360° horizontally and 180° vertically. In VR 360 live broadcasting, a user uploads live video streams to social media, e.g., Facebook and YouTube, using 360° cameras, and in the meanwhile the video is watched by other online users with VR headsets.

In this paper, we focus on the ultimate VR 360 that satisfies human eye fidelity. An ultimate VR video is encoded with 64 Pixels Per Degree (PPD), 12-bit color-depth and 120 Frames Per Second (FPS). The corresponding video bit rate is 2.3 Tbps. A 350:1 video compression ratio is expected to achieve with H.266 in a few years to come. This reduces the data rate of streaming an ultimate VR 360 video to 6.6 Gbps [1]. A further data rate reduction of VR 360 viewing can be realized by leveraging part-view transmission (120°×120° instead of 360°×180°), resulting in 1.5 Gbps data rate. Consequently, the ultimate VR 360 video requires networking infrastructure that supports 1.5 Gbps downlink for viewing and 6.6 Gbps uplink for live broadcasting, with Round-Trip Time (RTT) of less than 8.3 ms (frame interval of 120 FPS).

Home network access support for VR 360 is imperative because VR 360 videos are mostly used at home [1]. There are several challenges in building a home access network for the ultimate VR 360, including very asymmetric downlink and uplink demands, multiple users support and user viewing behavior. In addition, wireless access to VR 360 services is preferred over wire-line transmission. The wire-line connections to VR devices not only degrade user experience because the user cannot move freely with a cable connected headset, but also create a tripping hazard since the headset covers the user's eyes.

Considering the demanding networking requirements, e.g., data rate and latency, of the ultimate VR 360, we explore whether the most advanced current wireless technologies from both cellular communications and WiFi communications can support it. Specifically, we consider 5G in cellular communications, IEEE 802.11ac (operating in 5 GHz) and IEEE 802.11ad (operating in 60 GHz) in WiFi communications. For each selected technology, we survey its deployment status and technology features, as well as the performance determined by the standard specifications and/or empirically measured.

Through this preliminary exploration, we find that: (1) 5G is promised to provide 1 Gbps user experienced data rate and 1 ms latency. However, the data rate is lower than the ultimate VR 360 videos (both viewing and live broadcasting). (2) IEEE 802.11ac provides theoretical 6.9 Gpbs in the Access Point (AP) side and 3.5 Gbps in the client side and relatively low latency (2.3 ms RTT in our measurement). Therefore, it could support a single-user ultimate VR 360 viewing but fails to support the live broadcasting. (3) IEEE 802.11ad provides theoretical 6.8 Gbps in the AP side and 4.6 Gbps in the client side, but it incurs very long latency (62.7 ms in our measurement). With implementation enhancement, e.g., reducing the delay of antenna and beam tracking, IEEE 802.11ad could support a single-user ultimate VR viewing. Our preliminary results indicate that the ultimate VR 360 calls for more advanced wireless technologies that substantially boost networking performance (data rate and latency). The ongoing efforts are 5G extensions, IEEE 802.11ax (extension to IEEE 802.11ac) and IEEE 802.11ay (extension to IEEE 802.11ad) [3].

## II. VR 360 Video Requirements

The industry proposes different stages of VR 360 products, including the early stage, the entry-level experience stage, the advanced experience stage, and the ultimate experience stage [1]. These stages are different with regards to PPD, color depths and frame rates. (1) PPD is related to the display resolution. Given the viewing distance from a display, PPD and the widely used Pixels Per Inch (PPI) are transferable. PPD is independent of the viewing distance, and thus is favored in evaluating VR systems. For a user's FoV of $H$ horizontal degrees by $V$ vertical degrees, the length of the horizontal pixels and the vertical pixels is $H \times$PPD pixels and $V \times$PPD pixels respectively, with the display resolution of ($H \times$PPD) × ($V \times$PPD) pixels. It is commonly believed that human fovea can detect 60 PPD [4]. Therefore, in the proposed ultimate VR 360, 64 PPD is used to safeguard possible perception of discontinuous pixels. (2) Color depth is the number of bits to represent each red, green and blue component. For a system with color depth of $x$, the number of colors that the system can render is $2^{3x}$ considering the combination of the three-color components. For example, a video with 8-bit color depth can represent 16.8 million colors, whereas a video with 12-bit color depth can represent 4096 times more. HDMI 1.3 specification defines Deep Color, e.g., 12-bit color depth, in order to "eliminate any potential color banding artifacts that could be seen when there are not enough colors to properly display certain images" [5]. (3) Frame rate is the frequency at which consecutive images (video frames) appear on a display. A high frame rate is required to avoid motion blur. The perception of motion blur depends on many factors such as contrast, brightness, spatial factors, image content. It also varies among different human beings. 120 FPS is gaining popularity as witnessed by the support of AOMedia Video 1 (AV1, successor to VP9) and High Efficiency Video Coding (HEVC, successor to AVC).

The four stages of VR 360 products are proposed based on the forecast product evolution. The comparison of the data rate among different stages is illustrated in Figure 1. The early stage supports 11 PPD, equivalent to 3960×1980 pixels for a full-view frame of each eye; each pixel is encoded by 24 bits (color depth of 8 bit) and the frame rate is 30 PS. Correspondingly, the video bit rate of the early stage VR 360 is 11.3 Gbps (two eyes). As of early 2019, only the early stage VR 360 is fully supported. For example, HTC Vive Pro that was released in April 2018 renders 1600×1400 single-eye view for 110°×100° FoV, corresponding to 14 PPD. From Figure 1, we can clearly see that the early stage VR is toy-ish compared to the ultimate VR. In this paper, we focus on the ultimate VR 360 that satisfies human eye fidelity. The ultimate VR 360 adopts 64 PPD, 12-bit color depth, and 120 FPS, which results in video bit rate of 2.3 Tbps.

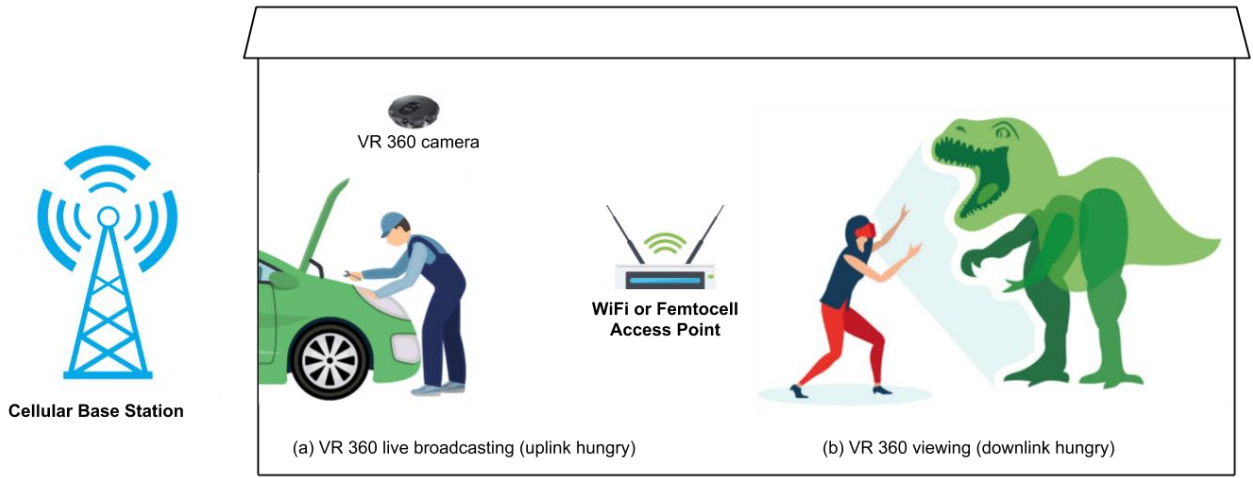

Fig. 2. Home application of VR 360, including uplink-hungry VR 360 live broadcasting and downlink-hungry VR 360 viewing. The VR cameras and the VR headsets are connected to the VR 360 servers via wireless connections such as cellular and WiFi

In addition to the extremely high data rate, VR 360 has stricter requirement for networking latency compared to traditional video streaming. In traditional video streaming, the whole video frames are transmitted, allowing buffering to alleviate network latency and variation. In VR 360 viewing, however, only the content within user's FoV are transmitted to the headset, in order to reduce the data rate. Compared to full-view transmission, the part-view transmission can adopt $120^o \times 120^o$ views, which reduces the data rate by 4.5 times. This corresponds to data rate reduction in Gbps levels. Therefore, part-view transmission is the *de facto* practice nowadays for the VR 360 viewing. However, compared to the full-view transmission, part-view transmission needs to track user head movement and promptly update the region of video content for transmission. Therefore, the RTT from the generation of head movement tracking to the rendering of FoV views on headset should be smaller than the interval of frame refresh. This requires network RTT to be smaller than 8.3 ms (120 FPS).

In practice, video compression techniques are used to greatly reduce the data rate requirement. Compared to traditional video streaming, VR 360 videos have higher compression ratio because the left-eye view and the right-eye view are partly overlapped (binocular overlap). Current encoding scheme HEVC (i.e., H.265) is expected to achieve 215:1 compression ratio for VR 360 [1]. The next-generation encoding scheme, called Future Video Codec (FVC or H.266) is estimated to be ready by 2021 with the goal of 50% more compression over HEVC. Therefore, it is reasonable to predict that in the next few years the compression ratio of 350:1 for VR 360 can be realized [1]. With H.266, the data rate of viewing the ultimate VR 360 (part-view transmission) can be reduced to 1.5 Gbps, and the data rate of the ultimate VR 360 live broadcasting (full-view transmission) can be reduced to 6.6 Gbps.

In summary, a network infrastructure for a single-user ultimate VR 360 requires supporting data rate of 1.5 Gbps downlink for viewing and 6.6 Gbps uplink for live broadcasting, with RTT of less than 8.3 ms.

## III. Challenges of Supporting VR 360 at Home

Home network access support for VR 360 is imperative because VR 360 entertainment videos are generally used at home [1]. Figure 2 depicts home applications of VR 360, including downlink-hungry viewing and uplink-hungry live broadcasting. The VR 360-degree cameras and the VR headsets are connected to VR 360 servers through wireless networks, e.g., via cellular base stations, femtocell access points or WiFi access points. It is challenging to build networking infrastructure to support the ultimate VR 360 at home because of many factors.

*Asymmetric Downlink and Uplink.* VR 360 viewing and VR 360 live broadcasting have extremely different traffic patterns on downlink and uplink. In VR 360 viewing, the traffic dramatically skews towards downlink like traditional video watching. Although VR 360 viewing needs to continuously track user's head movement and upload the tracking information to the server, the generated uplink traffic in Kbps, is negligible to the downlink traffic that is in Gbps level. On the other hand, VR 360 live broadcasting extremely skews towards uplink because it uploads captured video frames from a VR 360 camera to VR content media platform. Therefore, to carry the traffic of both the ultimate VR 360 viewing and live broadcasting, a network is required to support 1.5 Gbps downlink and 6.6 Gbps uplink. With the increasing popularity of online social media sharing, the landscape of home Internet access infrastructure may undergo a radical change: uplink speed is several times higher than downlink speed, in order to support VR 360 live broadcasting.

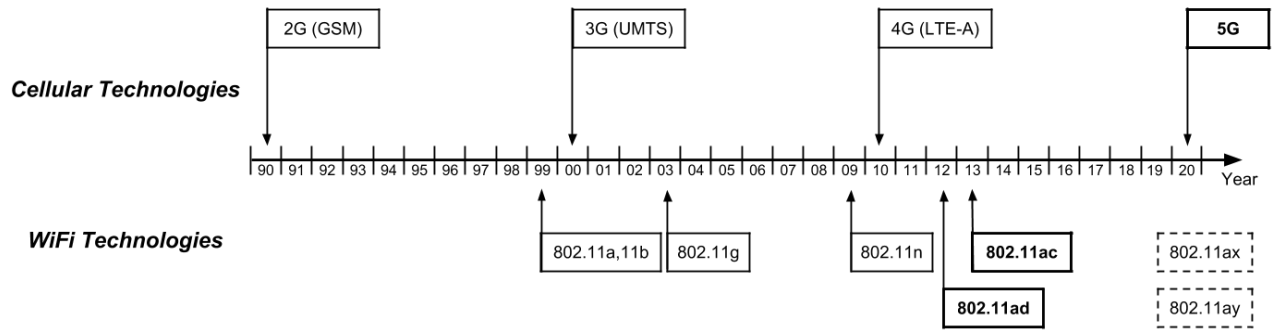

Fig. 3. Timeline for wireless technologies from cellular communications and WiFi communications. We explore 5G, IEEE 802.11ac and IEEE 802.11ad to determine to what extent they support the ultimate VR 360

*Multiple Users at Home.* Home access networks are required to support multiple users at home, in which each user may transmit/receive different VR 360 streams. Users in a household are usually connected to a single access point, e.g., WiFi or femtocell. In the past decade, the average number of people per household in the United States is about 2.5 [6]. The number is expected to remain stable in the next decades. The home access network, therefore, is preferred to simultaneously support three VR 360 streams. The resultant data rate of a typical household with three people can be 4.5 Gbps downlink if they are all watching VR 360 videos and astonishing 19.8 Gbps if they are all live broadcasting. In addition to the extremely high data rate, networks have difficulty in meeting the network RTT requirement, since higher data rate generally causes higher network RTT.

*VR 360 User Viewing Behavior.* According to the VR user behavior report [7] that was published in 2016, for users who bought VR devices in the past year, they spent an average time of 10 minutes in matching VR 360 video every day. It is expected that the viewing time will increase to 53 minutes in 2020 [1]. Similar to watching traditional videos, users mostly watch VR 360 video during the leisure time (19:00-23:00). The usage time ratio between the leisure time period and other time period is 8:2, and the ratio is expected to remain unchanged by 2025 [1]. The concentrated usage means that the home network traffic will increase tremendously during the leisure time compared to other time periods because of VR 360 applications. It remains a problem that how a network carrier makes profit from building extremely high-speed home network infrastructure while the usage time is concentrated. Differentiated pricing for different time periods of networking is still an effective strategy.

*Wireless Network Access.* Untethered high-quality VR 360 experience is highly desirable but challenging. Wireless technologies have limited throughput and high latency compared to wired-line technologies. For example, we calculate the average RTT of LTE and WiFi profiles from a public dataset [8] that was crowd-sourced in 16 countries. The results show that the average RTT is 246.1 ms and 86.5 ms for LTE and WiFi respectively, which is much larger than the target of 8.3 ms in the ultimate VR 360. With regards to the data rate, LTE networks support theoretical maximum of 326 Mbps and IEEE 802.11n provides theoretical maximum of 600 Mbps, which are also far from the data rate requirement of the ultimate VR 360.

Considering the demanding networking requirements of the ultimate VR 360, we explore whether the most advanced wireless technologies from both cellular communications and WiFi communications can support the ultimate VR 360. Figure 3 shows the timeline for cellular technologies and WiFi technologies. For cellular communications, we consider 5G that is about to be standardized and deployed. For WiFi technologies, we consider IEEE 802.11ac and IEEE 802.11ad.

### IV. Wireless Access: 5G

The fifth-generation cellular communication, i.e., 5G, is expected to be standardized by the early 2020s. Its design target is to achieve 10-100x peak data rate, 1000x network capacity, 10x energy efficiency, and 10-30x lower latency compared to its predecessor of 4G. Because of the exciting performance that 5G promises, the industry has already been building 5G ecosystems. For example, Huawei conducted 5G trials that show Gbps single-user downlink peak data rate with 2.6 GHz frequency bands, and launched 5G base station chips in January, 2019; AT&T plans to deploy 5G networks in 19 cities in early 2019; Qualcomm unveiled its 5G millimeter-wave module for smartphones in July 2018, and 5G smartphones (e.g., Samsung Galaxy S10 and Huawei Mate 30) will come out in 2019.

### A. 5G Technology Features

The data rate improvement of 5G over 4G mainly thanks to the following three technology categories: extreme densification and offloading, millimeter wave, and massive Multiple-Input and Multiple-Output (MIMO) [9].

*Extreme Densification and Offloading*. Making the cells smaller is straightforward but extremely effective to increase the network capacity. It has numerous benefits, including the reuse of spectrum in a given area, and the reduced number of users competing for resources at each cellular Base Station (BS). In principle, cells can shrink almost indefinitely without decreasing Signal-to-Interference Ratio (SIR) of users until every BS serves a single user, as long as the power-law pathloss models holds [10]. This allows each BS to devote its resource to a small number of users.

*Millimeter Wave*. 5G mmWave focuses on 30-100 GHz because of the mass market semiconductor technology constraints. mmWave was deemed unsuitable for wireless communications because of hostile propagation qualities. However, mmWave technology is maturing to combat the propagation issues and the hardware costs are falling. In 5G mmWave communication, dual connectivity will be an important feature to prevent loss of coverage, because lower frequency provides better coverage. Therefore, mmWave can be used for data transmission from small cells while the control plane operates at microwaves from macro cells. The dual connectivity ensures stable and reliable connections, while providing fast data transmission.

*Massive MIMO*. A massive MIMO system is typically defined as a system that uses a large number, i.e. 100 or more, of individually controllable antennas. It exploits the high spatial Degrees of Freedom (DoF) provided by the large number of antennas to realize spatial multiplexing, in which data are transmitted to several users on the same time-frequency resource. In addition, a massive MIMO system focuses the signal towards the intended receivers and thus minimizes intra-cell and inter-cell interference. Massive MIMO has the promise to provide a substantially increased spectral efficiency.

### B. 5G System Performance

Although 5G has not been standardized, it is expected to meet the IMT-2020 specifications [11]. IMT-2020 envisions different user experienced data rates to cover a variety of environments. For wide area coverage cases, such as urban areas, a user experience data rate of 100 Mbps is supported, whereas the hotspot cases such as indoor will support a user experience data rate of 1 Gbps. IMT-2020 would also provide 1 ms over-the-air latency. Therefore, it is reasonable to assume that 5G will provide 1 Gbps user data rate and 1 ms RTT.

## V. Wireless Access: IEEE 802.11ac

IEEE 802.11ac was released in 2013 (Figure 3). It is an evolution from IEEE 802.11n. The first AC router (Netgear R6300) was released in 2012 and the first AC-complaint smartphone (Samsung Mega) came out in 2013. Afterwards, more and more smartphones are equipped with AC WiFi modules. Despite having been on markets for several years, existing AC modules are far from the full capacity of IEEE 802.11ac standard. For example, the specification allows 8 spatial streams, while current AC devices can only support up to 4 spatial streams.

### A. IEEE 802.11ac Technology Features

IEEE 802.11ac improves IEEE 802.11n in many aspects, including wider channels, 256-QAM support, simplified beamforming, more spatial streams and multi-user MIMO (MU-MIMO) [12].

*Wider Channels*. IEEE 802.11ac supports 80 MHz and 160 MHz channels, in addition to 20 MHz and 40 MHz that are supported by IEEE 802.11n. IEEE 802.11ac supports two forms of 160 MHz channels: a single contiguous 160 MHz channel or an "80+80" channel that combine two 80 MHz channels and has the same capacity.

*256-QAM Support*. IEEE 802.11ac supports BPSK, QPSK, 16-QAM, 64-QAM and 256-QAM, with the latter transmits more bits per symbol while less robust to interference. The new introduction of 256-QAM enables IEEE 802.11ac to support 8 bits per symbol.

*Simplified Beamforming*. IEEE 802.11n implements beamforming between two devices by negotiating mutually agreeable beamforming functions from the menu of options. Very few vendors implement the same options, and thus there is almost no cross-vendor beamforming compatibility in IEEE 802.11n. Instead, IEEE 802.11ac radically simplifies the beamforming specifications to one preferred technical method.

*More Spatial Streams*. IEEE 802.11ac increases the maximum number of spatial streams from 4 to 8 at the AP, while the client side remains up to 4 spatial streams compared to IEEE 802.11n. The extra spatial streams of an AP can be used to transmit to multiple clients at the same time. To separate transmission among multiple users, the AP uses beamforming to focus each of the transmissions towards its respective client.

### B. IEEE 802.11ac System Performance

Once the channel bandwidth, Modulation and Coding Scheme (MCS), Number of Spatial Streams (NSS) and Guard Interval (GI) are determined, the data rate of an AC device can be found by looking up the data rate table (e.g., data rate table in [13]). A full-fledged AC AP supports a 160 MHz channel, MCS of 9, NSS of 8, and short

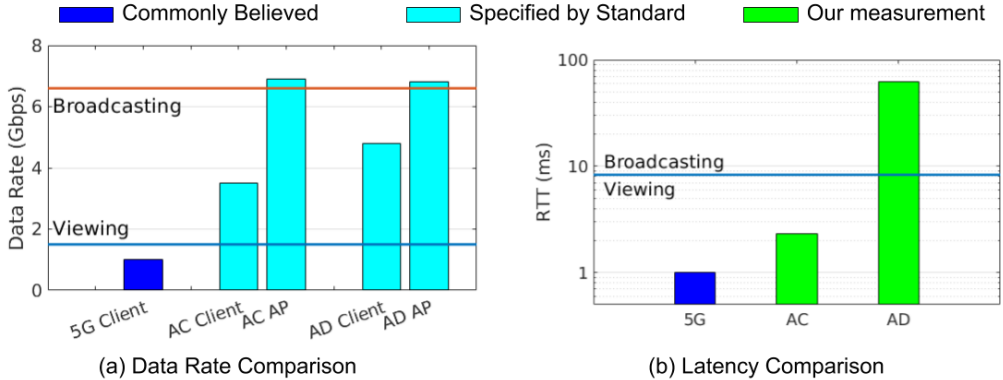

Fig. 4. Performance of 5G, IEEE 802.11ac and IEEE 802.11ad versus ultimate VR 360 requirements

GI, and the corresponding data rate is 6.9 Gbps. The capacity of a full-fledged AC client only differs from an AC AP with regards to the number of spatial streams. An AC client supports 4 spatial streams (NSS=4), with data rate of 3.5 Gpbs. To roughly quantify AC network RTT, we use two PC Engine APU 2 boards with AC modules (model: WLE650V5-18) that are closely placed, and ping each other, which gives an average RTT of 2.3 ms.

### VI. Wireless Access: IEEE 802.11ad

IEEE 802.11ad was standardized in 2012 (Figure 3). The first AD router (TP-link AD7200) was released in May 2016 and the first AD-complaint smartphone (ASUS Zenfone 4 Pro) came out in September 2017. The market adoption of IEEE 802.11ad is not successful. Very few smartphones have AD modules. However, as the first 802.11 standard on millimeter wave communication, IEEE 802.11ad was specifically designed for Gbps networking.

#### A. IEEE 802.11ad Technology Features

IEEE 802.11ad differs from legacy WiFi in many aspects including unique channel propagation behavior, novel beam training, and hybrid MAC channel access [14].

*Unique Channel Propagation Behavior.* IEEE 802.11ad operates in the 60 GHz frequency band, with up to 7 GHz unlicensed bandwidth. A typical coverage range of IEEE 802.11ad is within 10 m due to the high signal attenuation. Obstructions such as furniture and human body, can easily break the communication link of 60 GHz. Therefore, IEEE 802.11ad is more suitable to in-room environments where Line-Of-Sight (LOS) path is available or enough reflectors are present.

*Novel Beam Training.* IEEE 802.11ad introduces the concept of "virtual" antenna sectors that discretizes the antenna azimuth. Beamforming training between two devices happens at two different time: before their association when the direction between the two devices is unknown and during the data transmission interval. In addition, the beamforming protocol supports a training procedure for low antenna gain devices and can upload training parameters to a central network coordinator for channel access scheduling.

*Hybrid MAC Channel Access.* IEEE 802.11ad is intended to support various applications such as wireless PC display that requires real-time uncompressed video streaming, and bulk-file downloading that requires very high data rate. In contrast to legacy WiFi, IEEE 802.11ad adopts a hybrid MAC approach to address different application requirements. Specifically, IEEE 802.11ad incorporates three MAC scheduling: contention-based access, scheduled channel time allocation, and dynamic channel time allocation.

#### B. IEEE 802.11ad System Performance

The standard of IEEE 802.11ad specifies data rates that are supported (refer to the data rate table in [15]). For example, an AD AP achieves its maximum data rate using Orthogonal Frequency-Division Multiplexing (OFDM), 64-QAM, 13/16 code rate, 6 coded bits per single carrier ($N_{BPSC}$), 2016 coded bits per symbol ($N_{CBPS}$), 1638 data bits per symbol ($N_{DBPS}$). The corresponding data rate of an AD AP is 6.8 Gbps. An AD client adopts more energy-efficient transmission of Single Carrier (SC) rather than OFDM and achieves the maximum data rate of 4.6 Gbps when π/2-16QAM, 4 $N_{CBPS}$, and 3/4 code rate are used. To roughly quantify AD network RTT, we ping an AD router (Netgear Nightnhawk X10) from a closely located AD laptop (Acer TravelMate P), which gives an average RTT of 62.7 ms. The measured high RTT is related to the hardware we used, which might not be applicable to other devices (e.g., DisplayLink that is a 60 GHz wireless adapter for Oculus Rift). However, it is commonly observed that current 60 GHz implementations suffer from long delay due to antenna and beam tracking.

# VII. Preliminary Results and Concluding Remarks

Figure 4(a) shows the data rate supported by 5G, AC and AD, and the data rate requirement of the ultimate VR 360 viewing and live broadcasting. 5G does not provide enough speed for the ultimate VR 360. IEEE 802.11ac and IEEE 802.11ad have similar data rate support at AP side and client side: both support the ultimate VR 360 viewing but fail to work when it comes to the ultimate VR 360 live broadcasting because the clients cannot upload at an enough throughput. Figure 4(b) shows the corresponding RTTs of 5G, AC, and AD, and compares it with the latency requirement of the ultimate VR 360. 5G and AC offer acceptable RTTs; the current implementation of IEEE 802.11ad, however, incurs very high latency, probably due to the delays from beam tracking and alignment.

In summary, only IEEE 802.11ac supports the ultimate VR 360 viewing. With further implementation optimization, IEEE 802.11ad also has the potential to support it. However, none of them can support a single-user ultimate VR 360 live broadcasting. None of the existing wireless technologies can fully support multiple ultimate VR 360 users at home. Our findings highlight the need for more advanced wireless technologies.

Huanle Zhang (dtczhang@ucdavis.edu) is currently a Ph.D. student with the Department of Computer Science, University of California, Davis. He had internship at Simula Research Laboratory, Norway in 2017, Microsoft Research Asia, China and Bell Labs, USA in 2018. His research interests include virtual reality, augmented reality, mobile systems and applications, Internet of things, and artificial intelligence. He is a student member of the IEEE.

Ahmed Elmokashfi (ahmed@simula.no) is a Senior Research Scientist at Simula Metropolitan Centre for Digital Engineering and Simula Research Laboratory in Norway. He is currently heading the Centre for Resilient Networks and Applications (CRNA). His research interest lies in network Measurements and Performance. Over the past few years, he has been leading and contributing to the development, operation and management the NorNet testbed infrastructure, which is a countrywide measurement setup for monitoring the performance of mobile broadband networks in Norway.

Zhicheng Yang (zcyang@ucdavis.edu) is a PhD candidate in Computer Science at University of California, Davis, CA, USA. His current research interests include wireless sensing, 60 GHz mmWave communications and networks, and mobile computing related to health-care and smart agriculture. He is a student member of the IEEE.

Prasant Mohapatra (pmohapatra@ucdavis.edu) is serving as the Vice Chancellor for Research at University of California, Davis. He is also a Professor in the Department of Computer Science and served as the Dean and Vice-Provost of Graduate Studies at University of California, Davis. He was the Editor-in-Chief of the IEEE Transactions on Mobile Computing. He has served on the editorial board of the IEEE Transactions on Computers, IEEE Transactions on Mobile Computing, IEEE Transaction on Parallel and Distributed Systems, ACM WINET, and Ad Hoc Networks. He is a Fellow of the IEEE and a Fellow of AAAS.